\documentstyle[Sprocl]{article}
\begin{document}
\def\lambdabar{{\mathchar'26\mkern-9mu\lambda}}

\title{RELATIVITY AND NONLOCALITY}
\author{BAHRAM MASHHOON}
\address{Department of Physics and Astronomy\\
University of Missouri-Columbia\\
Columbia, Missouri 65211, USA\\
E-mail:  mashhoonb@missouri.edu}
\maketitle

\abstracts{The basic physical structure of the relativistic theory of
gravitation is discussed.  The significant role that the Hypothesis of Locality plays in
relativity theory is elucidated via the phenomenon of spin-rotation coupling.  The limitations
of this hypothesis are critically examined.  A nonlocal theory of accelerated observers is
presented and some of its observational consequences are described.}

\section{Introduction}
The general theory of relativity is in good agreement with experimental data available at
present; however, a microphysical understanding of gravity is still missing.  There are many
different attempts to reconcile general relativity with the quantum theory.  The approach
adopted here seeks to examine the physical foundations of general relativity in the light of
the principles of quantum physics.

To arrive at the basic physical postulates of general relativity, we adopt a
measurement-theoretic approach.  Therefore, we start with a global inertial frame of reference
in Minkowski spacetime with coordinates x$^{\mu} = (ct, {\bf x})$ and define the standard
observers to be the class of inertial observers at rest in this system.  Each standard observer
carries an orthonormal tetrad $\lambda^{\mu}\,_{(\alpha)}$, where
$\lambda^{\mu}\,_{(0)} = dx^{\mu}/d\tau$ is the vector tangent to the worldline of the observer
and
$\lambda^{\mu}\,_{(i)}\;\;,\;\; i = 1,2,3,$ are the spatial axes of the observer.  Hence
$\lambda^{\mu}\,_{(\alpha)} =
\delta^{\mu}\,_{\alpha}$ in this case.  The electromagnetic field as measured by the standard
observers is the projection of the Faraday tensor $F_{\mu\nu}$ on the tetrad of the observers,
\begin{equation}
F_{(\alpha)(\beta)} = F_{\mu\nu}\;\lambda^{\mu}_{(\alpha)}\;\lambda^{\nu}_{(\beta)}\;\;.
\end{equation}
Thus the electric and magnetic fields, $F_{(\alpha)(\beta)}\rightarrow({\bf E},
{\bf B})$, that appear in the standard form of Maxwell's equations are the fields as measured
by the standard inertial observers.  Imagine now an inertial observer moving uniformly with
speed $v$ along the $z$-axis.  To find the fields measured by this observer, one can either apply
the Lorentz transformation to the Faraday tensor, i.e. transform to the rest frame of the
observer, or to the tetrad frame and then project the field on the tetrad frame of the moving
observer in accordance with (1).  We adopt the second approach for the sake of convenience and
note that for this observer
\begin{eqnarray}
\lambda^{\mu}\,_{(0)} & = & \gamma(1, 0, 0, \beta)\;\;,\;\;\lambda^{\mu}\,_{(1)} = (0, 1, 0,
0),\nonumber\\
\lambda^{\mu}\,_{(2)} & = & (0, 0, 1, 0)\;\;\;\,\;,\;\;\lambda^{\mu}\,_{(3)} = \gamma(\beta, 0,
0, 1)\;,
\end{eqnarray}
where $\gamma$ is the Lorentz factor and $\beta = v/c$.  In this way Lorentz
invariance (or Poincar\'{e} invariance) provides a complete description of phenomena according
to inertial observers, since the fundamental laws of physics have all been formulated with
respect to such ideal observers.

To describe the measurements of arbitrary observers in Minkowski spacetime, we need to remove
two limitations that exist in our treatment thus far.  Let us first demand that inertial
observers be able to use arbitrary coordinates in Minkowski spacetime just as one uses
curvilinear coordinates in Euclidean space.  A detailed examination shows that this
limitation is purely mathematical and using the elegant tools of tensor calculus one can
simply extend the Lorentz-invariant theory of inertial observers to arbitrary admissible
coordinates in Minkowski spacetime.  We must next remove the restriction that thus far only
inertial observers make measurements; that is, we need a prescription for the results of
measurements by arbitrary accelerated observers in Minkowski spacetime.  To this end, a basic
hypothesis is needed to connect the measurements of accelerated observers to those of inertial
observers.  It is natural to expect such a connection, since the whole observational basis of
Lorentz invariance involves measurements made by accelerated observers such as those on the
rotating Earth.  The assumption that is employed in the theory of relativity is the Hypothesis
of Locality, which postulates the local equivalence of an accelerated observer with an
instantaneously comoving (hypothetical) inertial observer.  The worldline of an accelerated
observer is a curved path in Minkowski spacetime; therefore, the hypothesis of locality
replaces the curve by the line tangent to the curve at each event.  This hypothesis is
embodied in the assumption that clocks and rods are locally unaffected by acceleration [1].  

The hypothesis of locality is clearly valid in the Newtonian mechanics of point masses, since
the accelerated and inertial particles instantaneously share the same {\it state}
$({\bf x, v})$; therefore, no extra physical assumption is needed in the discussion of inertial
effects in Newtonian mechanics.  In relativity theory, however, the pointwise character
of the hypothesis of locality imposes severe limitations on the notion of distance [2].
Moreover, the hypothesis of locality implies that an accelerated observer in Minkowski spacetime
is also endowed with an orthonormal tetrad frame
$\lambda^{\mu}\,_{(\alpha)}(\tau)$ that instantaneously coincides with that of the comoving
inertial observer.  From
\begin{equation}
\frac{D\lambda^{\mu}{}_{(\alpha)}}{d\tau} = \phi_{\alpha}{}^{\beta}\,\lambda^{\mu}{}_{(\beta)}
\end{equation}
and the decomposition of $\phi_{\alpha\beta} = -\phi_{\beta\alpha}$ in terms of
translational acceleration ${\bf g}$ and rotational frequency of the spatial frame
$\bf{\Omega}$ with respect to nonrotating spatial axes, $\phi_{\alpha\beta}\rightarrow(\bf{g},
\bf{\Omega})$, we arrive at the acceleration lengths $c^2/g$ and $c/\Omega$ that characterize
the curvature of the worldline [2-4].  Neglecting the instantaneous curvature of the
worldline, as demanded by the hypothesis of locality, means that these acceleration lengths
are immeasurably large compared to any other relevant scale in the physical phenomenon under
consideration.  This limitation will be discussed in the following section. 

The special theory of relativity is therefore based upon Lorentz invariance and the hypothesis
of locality.  To explain the measurements of an observer in a gravitational field, Einstein's
principle of equivalence is indispensable.  According to this heuristic principle, an observer
in the gravitational field is locally equivalent to a certain accelerated observer in Minkowski
spacetime.  In conjunction with the hypothesis of locality, Einstein's principle of
equivalence implies that every observer in a gravitational field is locally inertial.  The
simplest way to connect the local inertial frames of observers in a gravitational field is via
the Riemannian curvature of the spacetime manifold.  This curvature is then identified with
the gravitational field in Einstein's theory, though many extensions and generalizations are
possible.  Free test particles and null rays thus follow geodesics of the curved spacetime in
general relativity.  It remains to give the field equations for gravitation; in general
relativity, this last step involves the simplest generalization of Newtonian gravitation that
is consistent with the spacetime structure.  The physical basis of general relativity thus
consists of [5]
\\
\indent (1)  Lorentz Invariance,\\
\indent (2)  Hypothesis of Locality,\\
\indent (3)  Einstein's Principle of Equivalence,\\
\indent (4)  Correspondence with Newtonian Gravitation; Field Equations.\\
The great success of relativistic quantum theory means that we must begin with the
hypothesis of locality in our quest for the integration of relativity theory with quantum
physics. 

\section{Spin-Rotation Coupling}
Classically, motion takes place via classical particles as well as electromagnetic waves.  The
latter have an intrinsic length scale characterized by their wavelength $\lambda$.  For most
laboratory experiments, however, $\lambda$ is very small compared to the usual acceleration
lengths, since $c^2/g\simeq 1$ lyr for the acceleration of gravity on the Earth and
$c/\Omega\simeq 28$ AU for the proper rotation frequency of the Earth.  To explore the
consequences of the hypothesis of locality explicitly, let us consider a simple thought
experiment.  Imagine an observer following a circle of radius $r$ about the origin in the ($x,
y$)-plane of a global inertial frame of reference such that the observer rotates uniformly
around the $z$-axis with frequency
$\Omega$.  A plane monochromatic electromagnetic wave propagates along the $z$-direction.  The
frequency of the wave according to the standard observers is $\omega$.  What is the frequency
as measured by the rotating observer?  Introducing the hypothetical comoving inertial observer
as in the hypothesis of locality, we can employ the transverse Doppler effect between the
instantaneous inertial frame and the global inertial frame, and conclude that the answer is
simply $\gamma\omega$ due to time dilation.  Here $\gamma = (1 - \beta^2)^{-1/2}$ and $\beta =
r\Omega/c$.  On the other hand, we can apply this hypothesis to the field and use the tetrad
frame of the observer,
\begin{eqnarray}
\lambda^{\mu}\,_{(0)} & = & \gamma(1, -\beta{\rm sin}\varphi\;\;,\;\;\beta{\rm cos}\varphi,
0)\;\;,\;\;\lambda^{\mu}\,_{(1)}=(0, {\rm cos}\varphi, {\rm sin}\varphi, 0),\nonumber\\
\lambda^{\mu}\,_{(2)} & = & \gamma(\beta, - {\rm sin}\varphi\;\;,\;\; {\rm cos}\varphi,
0)\;\;,\;\;\lambda^{\mu}\,_{(3)} = (0, 0, 0, 1)\;\;,
\end{eqnarray}
where $\varphi = \Omega t = \gamma\Omega\tau$, to find $F_{(\alpha)(\beta)}(\tau)$ in
accordance with (1) and then Fourier analyze this result --- a nonlocal operation --- to
conclude that the frequency measured by the observer is
\begin{equation}
\omega^{\prime} = \gamma(\omega\mp\Omega)\;\;.
\end{equation}
Here the upper sign is for positive helicity radiation (RCP or right circularly
polarized) and the lower sign is for negative helicity radiation (LCP or left circularly
polarized).  Equation (5) has a simple physical interpretation: in RCP (LCP) radiation, the
electromagnetic field rotates in the positive (negative) sense along the direction of
propagation and hence the observer sees a kind of rotational Doppler effect.  Multiplying both
sides of equation (5) by $\hbar$, we find $E^{\prime} = \gamma(E -
\mbox{\boldmath $\sigma$}\cdot{\bf{\Omega}})$, which indicates that the spin of the photon
couples to rotation.  This is an example of the general phenomenon of spin-rotation coupling
that is supported by experiment [6-8].

Several features of spin-rotation coupling (5) should be mentioned.  Writing (5) as
$\omega^{\prime} = \gamma\omega(1\mp\Omega/\omega)$, we note that the result is different from
$\gamma\omega$ that was obtained by the simple application of the hypothesis of locality by a
term of the form
$\Omega/\omega = \lambdabar/(c/\Omega)$, where $c/\Omega$ is the acceleration length.  Thus
$\omega^{\prime}=\gamma\omega$ for a null ray with
$\lambda = 0$; hence, the Doppler effect is in general strictly valid only in the geometric
optics limit [9].  Moreover, if the incident RCP and LCP waves have the same amplitude
according to the standard observers, then their amplitudes measured by the rotating observer
will also be the same.  Finally, equation (5) is an example of the more general formula
$\omega^{\prime} = \gamma(\omega - M\Omega)$ for oblique incidence, where $M$ is a total spin
parameter given by $M = 0, \pm 1, \pm 2, ...$, for a scalar or a vector field, while $M\mp 1/2
= 0, \pm 1, \pm 2, ...$, for a Dirac field.  Let us note that $\omega^{\prime} = 0$ for
$\Omega = \omega/M$, i.e. an observer can stand completely at rest with respect to the
radiation field by a mere rotation.  This is particularly clear in (5) for an observer that
rotates with frequency $\Omega = \omega$ with respect to an incident RCP wave.  This basic
difficulty, which does not arise in the Doppler effect, leads us to re-examine the physical
basis of the hypothesis of locality.  

\section{Nonlocality}
The hypothesis of locality is clearly an approximation in the same sense that a curve can be
locally approximated by its tangent line.  An accelerated observer passes through a
continuous infinity of momentarily comoving inertial observers; therefore, the measurements of
the accelerated observer could in general be related to the measurements of the whole
sequence of hypothetical inertial observers.  Let $\psi$ be the field according to the standard
inertial observers and $\Psi^{\prime} = \Lambda\psi$ be the field as measured by the
instantaneously comoving inertial observer.  Specialized to the electromagnetic case, $\psi$
and $\Psi^{\prime}$ are column 6-vectors representing $F_{\mu\nu}$ and $F_{(\alpha)(\beta)}$,
respectively, so that $\Lambda$ is a $6\times6$ matrix determined via equation (1).  The most
general relationship between the result of field measurement by an accelerated observer $\Psi$
and $\Psi^{\prime}$ that is consistent with causality and the superposition principle is 
\begin{equation}
\Psi(\tau) = \Psi^{\prime}(\tau) + \int_{\tau_{0}}^{\tau}\;K(\tau, x)\Psi^{\prime}(x)\;dx\;\;,
\end{equation}
where $\tau_0$ is the instant at which the acceleration is turned on and $K(x, y)$
is a kernel that must be determined on the basis of further physical hypotheses.  If $K = 0$,
then the hypothesis of locality is satisfied; therefore, we expect the integral part in (6)
that signifies the deviation from locality to be of order $\lambda/{\cal L}$, where $\lambda$
is the intrinsic scale of fluctuations of the field and ${\cal L}$ is an acceleration length
of the observer. 

The general theory of integral relationships of the form (6) was originally developed by Vito
Volterra [10], who showed that in the space of continuous functions the relationship between
$\Psi$ and $\Psi^{\prime}$ is unique.  This result has been essentially extended to the Hilbert
space of square-integrable functions by Tricomi [11].

Searching for a basic physical hypothesis that would determine the kernel $K$, we turn to the
phenomenon of spin-rotation coupling.  The formula (5), that is based on the hypothesis of
locality for the field, has the simple consequence that the RCP radiation field becomes
static $(\omega^{\prime} = 0)$ according to rotating observers for $\omega = \Omega$.  That
is, by a mere rotation an observer can stay completely at rest with a radiation field.  Let
us recall that if Maxwell's equations are assumed to hold in all inertial frames, then the
speed of light $c$ must be constant for all inertial observers.  To ensure this fact, no
inertial observer can move with speed $c$; therefore, in the relativistic Doppler formula,
$\omega^{\prime} = \gamma\omega(1 - {\bf v}\cdot \bf{\hat{k}}/c),\omega^{\prime}$
can never be zero.  For the motion of an inertial observer along the direction of propagation of
the wave,
$\omega^{\prime} = \omega(1 - \beta)^{1/2}/(1 + \beta)^{1/2}$ can be made as small as possible
for $\beta\rightarrow 1$, yet the mathematical limit is avoided due to the physical restriction
that $\beta < 1$.  The only case for which $\omega^{\prime} = 0$ is that $\omega = 0$, i.e.
if one inertial observer measures a time-independent field, then all inertial observers
measure time-independent fields.  In this way the existence of a quantum of radiation as well
as the number of quanta in the field acquires an observer-independent status.  It is
worthwhile to generalize this basic consequence of Lorentz invariance for inertial observers
to all observers.  Therefore, we demand that if $\Psi$ becomes constant, then $\psi$ should be
constant as well.  Let us note that this basic assumption is related to a natural
generalization of quantum mechanics for noninertial observers.  A postulate of classical
mechanics [12] is that an observer can be comoving with a particle, since they can have the
same position and velocity.  A classical observer cannot be comoving with a quantum particle,
however, as a consequence of Heisenberg's uncertainty principle.  The state of the quantum
particle in fact satisfies a wave equation and hence a classical observer cannot stay at rest
with respect to a fundamental wave.  

Before implementing the general requirement that no observer can stay completely at rest with
respect to a radiation field, we need to discuss the resolvent kernel for (6).  This is done
in the next section.  

\section{Resolvent Kernel}

Let us start with an integral equation of the form 
\begin{equation}
\phi(x) = f(x) + \lambda_{0}\int^{x}_{a}\;K(x, y)\;\phi(y)dy\;\;,
\end{equation}
where $\lambda_0$ is a constant parameter.  It can be shown that 
\begin{equation}
f(x) = \phi(x) + \lambda_{0}\int^{x}_{a}\;R(x, y)\;f(y)dy\;\;,
\end{equation}
where $R$ is the resolvent kernel [10, 11].  To find $R$ in terms of $K$, let us
define the successive iterated kernels of $K$ by $K_1(x, z) = K(x, z)$ and 
\begin{equation}
K_{n+1}(x, z) = \int^x_zK(x, y)K_n(y, z) dy\;\;.
\end{equation}
It can be shown [10,11] that
\begin{equation}
R(x, y) = -\sum^{\infty}_{n = 1}\lambda_0\,^{n-1}K_n(x, y)\;\;.
\end{equation}

Consider now the special case in which $K(x, y) = k(x - y)$, i.e. the kernel is of the
convolution (Faltung) type.  One can easily show, by letting $x - y = u$ and $x - z = t$ in (9),
that all the iterated kernels are of convolution type with 
\begin{equation}
k_{n+1}(t) = \int^t_0k(u)k_n(t - u) du\;\;,
\end{equation}
i.e. the iterated kernels can be obtained by successive convolutions of $k$ with
itself.  Denoting the convolution operation by a star,
\begin{equation}
\phi\ast\chi(t) = \int^t_0\phi(u)\chi(t-u)du\;\;\;\;=\chi\ast\phi(t)\;\;,
\end{equation}
and writing $\phi\ast\phi = \phi^{\ast 2}$, etc., we can express the resolvent kernel
$R(x, y) = r(x - y)$ as 
\begin{equation}
r(t) = -\sum^{\infty}_{n=1}\,\lambda_0\,^{n-1}\;k^{\ast n}(t)\;\;.
\end{equation}

Finally, let us note that if $K(x, y) = k_0(y)$, then the iterated kernels with $n>1$ and
the resolvent kernel are in general functions of both variables.  

\section{Kernel K}
The integral equation (6) may be written in the form 
\begin{equation}
\Psi(\tau) = \Lambda(\tau)\psi(\tau) + \int^{\tau}_{\tau_0}\;K(\tau,
\tau^{\prime})\Lambda(\tau^{\prime})\psi(\tau^{\prime})d\tau^{\prime}\;\;,
\end{equation}
such that at $\tau = \tau_0, \Psi(\tau_0) = \Lambda(\tau_0)\psi(\tau_0)$.  To
implement our basic hypothesis, we assume that $\Psi(\tau)$ is constant, i.e. $\Psi(\tau) =
\Psi(\tau_0)$, and expect that $\psi(\tau)$ will be constant as well, i.e. $\psi(\tau) =
\psi(\tau_0)$; hence, 
\begin{equation}
\Lambda(\tau_0) = \Lambda(\tau) + \int^{\tau}_{\tau_0}K(\tau,
\tau^{\prime})\Lambda(\tau^{\prime})d\tau^{\prime}\;\;.
\end{equation}
This is the basic integral equation for the determination of $K$.  Using the notion of
the resolvent kernel discussed in the previous section, we can write
\begin{equation}
\Lambda(\tau) = \Lambda(\tau_0) + \int^{\tau}_{\tau_0}R(\tau,
\tau^{\prime})\Lambda(\tau_0)d\tau^{\prime}\;\;,
\end{equation}
so that we have
\begin{equation}
\int^{\tau}_{\tau_0}R(\tau, \tau^{\prime})d\tau^{\prime} = \Lambda(\tau)\Lambda^{-1}(\tau_0) -
1\;\;.
\end{equation}
This integral relation, in which the right side is known, is not sufficient to
determine the resolvent kernel.  We need an extra assumption.  In this paper we explore two
possibilities regarding $K$: case (i) $K(\tau, \tau^{\prime}) = k(\tau -
\tau^{\prime})$ and case 
(ii) $K(\tau, \tau^{\prime}) = k_0(\tau^{\prime})$.

In the first case, a convolution-type kernel implies that $R(\tau, \tau^{\prime}) = r(\tau -
\tau^{\prime})$ and (17) becomes
\begin{equation}
\int^{\tau-\tau_0}_{0}r(u) du = \Lambda(\tau)\Lambda^{-1}(\tau_0) - 1
\end{equation}
and a simple differentiation results in 
\begin{equation}
r(x) = \frac{d\Lambda(x + \tau_0)}{dx}\Lambda^{-1}(\tau_0)\;\;.
\end{equation}
Once $r(x)$ is determined from (19), (13) with $\lambda_0 = -1$
implies, via the reciprocity between $K$ and $R$, that $K(\tau,
\tau^{\prime}) = k(\tau - \tau^{\prime})$ is given by 
\begin{equation}
k(x) = \sum^{\infty}_{n=1}(-1)^{n}r^{\ast n}(x)\;\;.
\end{equation}
In particular if $r(x) = 0$, then $k(x) = 0$ and the nonlocality disappears. 

In the second case, we have via differentiation of (15) that 
\begin{equation}
k_0(\tau) = -\frac{d\Lambda(\tau)}{d\tau}\Lambda^{-1}(\tau)\;\;.
\end{equation}

It is interesting to illustrate these results for the electrodynamics of linearly accelerated
systems.  Consider an obsever at rest on the $z$-axis for $-\infty<\tau<\tau_0$.  At $\tau =
\tau_0$, the observer accelerates linearly from rest along the $z$-axis with acceleration
$g(\tau)>0$.  Let 
\begin{equation}
\theta(\tau) = \int^{\tau}_{\tau_0}g(\tau^{\prime})d\tau^{\prime}\;\;,
\end{equation}
$C = \cosh\theta$, and $S = \sinh\theta$; then, the nonrotating orthonormal
tetrad along the observer worldline is given by 
\begin{eqnarray}
\lambda^{\mu}\,_{(0)} & = & (C, 0, 0, S)\;\;,\;\;\lambda^{\mu}\,_{(1)}\: = (0, 1, 0,
0)\;,\nonumber\\
\lambda^{\mu}\,_{(2)} & = & (0, 0, 1, 0)\;\;\;\,,\,\;\;\lambda^{\mu}\,_{(3)} = (S, 0, 0, C)\;.
\end{eqnarray}
Equation (1) can be written as $F^{\prime} = \Lambda F$, where $F$ is a column
6-vector with {\bf E} and {\bf B} as components and 
\begin{equation}
\Lambda = \left[ \begin{array}{cc} U & V\\ -V & U
\end{array} \right]\quad , 
\end{equation}
where
\begin{equation}
U = \left[ \begin{array}{ccc} C & 0 & 0\\
0 & C & 0\\
0 & 0 & 1
\end{array} \right]\quad , \;\;V = SI_3\;\;.
\end{equation}
Here $I_i, (I_i)_{jk} = -\epsilon_{ijk}$, is a $3\times 3$ matrix that is proportional
to the operator of infinitesimal rotations about the $x^i$-axis.  Let us note that
$\Lambda(\tau_0)$ is the identity matrix in this case.  

For the kernel of convolution type, (19) implies that 
\begin{equation}
r(\tau - \tau_0) = g(\tau) \left[ \begin{array}{cc} R_1 & R_2\\ -R_2 & R_1
\end{array} \right]\quad , 
\end{equation}
where $R_1 = SJ_3, R_2 = CI_3$, and $(J_k)_{ij} = \delta_{ij} -
\delta_{ik}\delta_{jk}$.  The convolution kernel
$k(x)$ constructed from
$r(x)$ as in (20) is rather complicated in general; however, it takes a very simple form for
{\it uniform} acceleration.  In fact, $k$ is in general a constant for arbitrary
{\it uniform} acceleration.  In the case under consideration, an explicit calculation
using (20) results in 
\begin{equation}
k = -g_0 \left[ \begin{array}{cc} 0 & I_3\\ -I_3 & 0
\end{array} \right]\quad , 
\end{equation}
where $g(\tau) = g_0$ is the constant magnitude of acceleration.  

On the other hand, for the case $K(\tau, \tau^{\prime}) = k_0(\tau^{\prime}), k_0$ can be
simply computed using (21) and the result is 
\begin{equation}
k_0(\tau) = -g(\tau)\left[ \begin{array}{cc} 0 & I_3\\ -I_3 & 0
\end{array}\right]\quad \;\;.
\end{equation}
Let us remark that for {\it uniform} acceleration, we have the same result as
in the convolution case (27); in fact, this is generally the case for arbitrary constant
acceleration.  That is, it turns out that the kernel $K$ is constant for uniform translational
and rotational accelerations and that this unique result has an interesting interpretation in
terms of the spacetime connection [13].  However, for nonuniform acceleration the two approaches
give different results.  

An important distinction between the two approaches is the nature of the memory of nonuniform
acceleration that lingers after the acceleration has been turned off, say, at $\tau_1 >
\tau_0$.  In case (i), the field as measured by the observer would still be nonlocal in general,
while in case (ii) the field would be local and the integral term in (14) would simply be a
{\it constant} field that can be easily eliminated by a constant recalibration of the
measuring devices carried by the observer.  Future experiments may use such differences to
distinguish between the two possibilities; that is, one could determine whether the kernel $K$
is of convolution type as in case (i) or $K(\tau, \tau^{\prime}) = k_0(\tau^{\prime})$ as in
case (ii).  In this connection, it is interesting to note here that convolution-type kernels
have been employed in the phenomenological studies of material media for a long time.  Such
studies of history-dependent phenomena apparently began with Poisson's work [14]
and have continued in hysteresis theory and continuum physics to the present time [15].  For
this reason, a convolution-type kernel was assumed at the outset in the nonlocal theory of
accelerated systems presented in [16].  

Let us now discuss certain consequences of nonlocality that follow from the theory developed
here.  It follows from (19) and (21) that for constant $\Lambda$ the nonlocality disappears. 
This is the case for all fields if the observer is inertial.  Moreover, for a fundamental
scalar (or pseudoscalar) field $\Lambda = 1$ and hence nonlocality disappears.  Thus it would
be possible in principle for an observer to stay completely at rest with respect to a scalar
(or pseudoscalar) field.  This is forbidden by our postulate, however.  The theory developed
here therefore excludes the possibility of existence of a basic scalar (or pseudoscalar) field
in nature.  Only composite fields of this type can occur; hence, should the Higgs boson be
discovered, it would have to be a composite particle.  

In connection with the observational consequences of nonlocality, it is interesting to return
to the phenomenon of spin-rotation coupling discussed in section 2.  Dealing with the uniformly
rotating observer, the two cases considered in (19) and (21) result in the same kernel that has
been studied in some detail in [16].  A direct consequence of nonlocality in this case is that
the field amplitude as measured by the rotating observer depends on the helicity of the incident
radiation.  According to the hypothesis of locality, the measured amplitudes will be
independent of the helicity of perpendicularly incident plane wave, i.e. if the amplitudes of RCP
and LCP waves are equal according to the standard observers in the inertial frame, they will
also be equal according to the comoving inertial observers.  This has a direct analog in the
impulse approximation of quantum scattering theory [17].  On the other hand, nonlocality
implies that for $\omega > \Omega$ the amplitude will be larger by $(1 - \Omega/\omega)^{-1}$ for
RCP radiation and smaller by $(1 + \Omega/\omega)^{-1}$ for LCP radiation.  For microwaves of
$\lambda\simeq 1$ cm incident normally at an observer rotating with a frequency of 300 Hz,
$\Omega/\omega\simeq 10^{-8}$.

\section{Discussion}
It is important to subject the nonlocal theory developed here to direct experimental test.  For
this purpose one could re-examine the whole observational basis of the theory of relativity ---
since such data generally involve observers that are accelerated --- to search for evidence of
nonlocality.  In this connection, an important difficulty is that the electrodynamics of
accelerated media is rather poorly understood.  It would therefore be a complicated matter to
isolate the vacuum nonlocality under consideration here.  

On the other hand, if the idea of nonlocality has merit, then we expect that the nonlocal
theory would be in better agreement than the standard theory with quantum mechanics in the
correspondence limit.  For instance, one may consider the accelerated observer to be --- in a
certain approximate sense --- an electron in a Rydberg state.  By studying various transition
rates, one may be able to provide evidence for the nonlocal theory.  In particular, for the
uniformly rotating observer, nonlocality implies that there is a relative increase (decrease) in
the amplitude of the measured field by $\Omega/\omega(-\Omega/\omega)$ for
incident RCP(LCP) radiation with $\omega \gg\Omega$.  Preliminary efforts in this direction
indicate that this prediction of the nonlocal theory is in qualitative agreement with
quantum-mechanical results [18].  

\section*{Acknowledgments}
I am grateful to Friedrich Hehl and Yuri Obukhov for interesting discussions and
correspondence.  I thank the organizers of the Spanish Relativity Meeting (EREs2000) for their
kind invitation and excellent hospitality.

\end{document}